\documentclass[a4paper,11pt]{article}
\usepackage{pos}
\usepackage{braket}
\graphicspath{{plots/}}

\title{Tensor-network simulation of the strong-coupling $U(N)$ model}

\author*[a]{Pascal Milde}
\author[a]{Jacques Bloch}
\author[a,b]{Robert Lohmayer}

\affiliation[a]{Institute for Theoretical Physics, University of Regensburg, Regensburg, Germany}
\affiliation[b]{RCI Regensburg Center for Interventional Immunology, Regensburg, Germany}

\emailAdd{pascal.milde@stud.uni-regensburg.de}
\emailAdd{jacques.bloch@ur.de}
\emailAdd{robert.lohmayer@ur.de}

\abstract{We apply tensor network methods to study the strong-coupling $U(N)$ model in its dimer formulation. In three and four dimensions, we investigate the chiral condensate as a function of the quark mass and the degree of the symmetry group, and find good agreement with Monte Carlo simulations. Particularly interesting is the study of chiral symmetry breaking as a function of the mass and the volume, which clearly shows that this symmetry is spontaneously broken in the limit of infinite volume and zero mass.}

\FullConference{%
 The 38th International Symposium on Lattice Field Theory, LATTICE2021
  26th-30th July, 2021
  Zoom/Gather@Massachusetts Institute of Technology
}

\begin{document}
\maketitle

\section{Introduction}
We present first tensor-network results for the strong-coupling limit of $U(N)$ gauge theory in three and four dimensions. Using the higher-order tensor renormalization group (HOTRG) method we reproduce the Monte Carlo results obtained in the monomer-dimer representation by Rossi and Wolff using a Metropolis algorithm in 1984 \cite{Rossi:1984cv} and by Adams and Chandrasekharan using a directed path algorithm in 2003 \cite{Adams:2003cca}. For very small lattices we find agreement up to twelve digits with the exact analytical results for the chiral condensate. We are also able to verify dynamical chiral symmetry breaking (DCSB) at small masses when using large lattices, which are easily accessible in tensor network simulations. 

\section{Partition function}
The partition function of the strongly coupled $U(N)$ gauge theory is
\begin{align}
Z = \int D\Psi D\bar{\Psi} dU \exp (S(\Psi,\bar{\Psi},U)) ,
\end{align}
where the action $S$ only consists of the fermion action
\begin{align}
S = S_f
=\frac{1}{2}\sum_{x,\mu}\Gamma_{\mu}(x)[\bar{\Psi}(x)U_{\mu}(x)\Psi(x+\hat{\mu})-\bar{\Psi}(x+\hat{\mu})U^\dagger_{\mu}(x)\Psi(x)]
+ m\sum_x\bar{\Psi}(x)\Psi(x) 
\end{align}
as the gauge action is absent in the strong-coupling limit ($\beta=0$). 
In the fermion action, $\Psi(x)$ and $\bar{\Psi}(x)$ represent the $N$-dimensional fermion and anti-fermion fields, respectively, $m$ is the fermion mass, $U_{\mu}(x)$ are the gauge links in the fundamental representation of $U(N)$, and $\Gamma_{\mu}(x)$ denote the staggered phase factors which include a temperature parameter $\tau$ in the time direction, i.e., $\Gamma_{1}(x)=\tau$ and $\Gamma_\mu(x)=\exp[i\pi(x_1+\cdots+x_{\mu-1})]$, $\mu=2,3,\ldots,d$  \cite{Adams:2003cca}. 

In the strong-coupling limit, the gauge fields and the fermion fields can be integrated out, and the partition function can be represented by a dimer-monomer system, as was originally proposed by Rossi and Wolff (1984) \cite{Rossi:1984cv}. Using this dual formulation, the partition function can be written as a fully contracted tensor network,
\begin{align}
Z(m)=\sum_{\{k\}}\prod_{x}T_{k_{x,1},k_{x-\hat{1},1},\ldots,k_{x,d},k_{x-\hat{d},d}}
\label{Zm}
\end{align}
with local tensor 
\begin{align}
T_{k_{x,1},k_{x-\hat{1},1},...,k_{x-\hat{d},d}}=\left( \prod_\mu^d \sqrt{\alpha_{k_{x,\mu}}\alpha_{k_{x-\hat{\mu},\mu}}}\right) \frac{1}{2^{\sigma_x}}\frac{N!}{(N-\sigma_x)!}m^{N-\sigma_x}\tau^{{{k_{x,\hat{1}}}+{k_{x-\hat{1},\hat{1}}}}}\theta(N-\sigma_x) ,
\end{align}
where $k_{x,\mu}=0,\ldots,N$, $\sigma_x=\sum_{\mu=1}^d (k_{x,\mu}+k_{x-\hat{\mu},\mu})$, $\alpha_k=\frac{(N-k)!}{k!N!}$, $\sum_{\{k\}}$ is the sum over all configurations in $k$-space, and $\theta(x)$ is the Heaviside step function (defined as $\theta(x\geq0)=1, \text{else 0}$).

\begin{figure}[ht]
\scriptsize
\parbox{0.28\textwidth}{\hspace{16mm}{$T^{(0)} \to T^{(1)}$}}
\normalsize\phantom{$\to$}
\scriptsize
\parbox{0.16\textwidth}{\hspace{6mm}{$T^{(1)} \to T^{(2)}$}}
\normalsize\phantom{$\to$}
\scriptsize
\parbox{0.16\textwidth}{\hspace{6mm}{$T^{(2)} \to T^{(3)}$}}
\normalsize\phantom{$\to$}
\scriptsize
\parbox{0.16\textwidth}{\hspace{3mm}{$T^{(3)} \to T^{(4)}$}}
\normalsize\phantom{$\to$}
\scriptsize
\parbox{0.10\textwidth}{\hspace{-6mm}{$T^{(4)} \to Z$}}
\\[-2mm]

\normalsize
\parbox{0.28\textwidth}{\centering\includegraphics[scale=0.78]{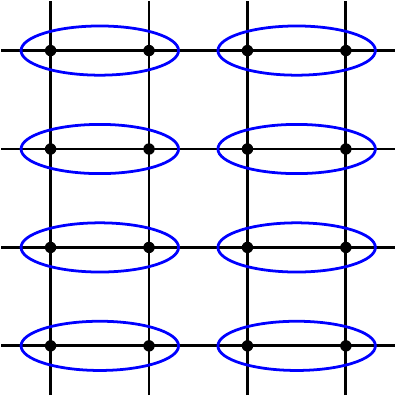}}
$\to$
\parbox{0.16\textwidth}{\centering\includegraphics[scale=0.78]{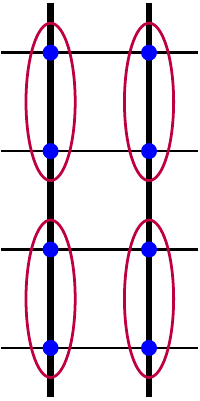}}
$\to$
\parbox{0.16\textwidth}{\centering\includegraphics[scale=0.78]{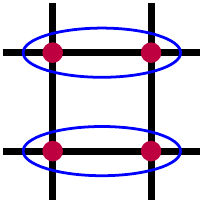}}
$\to$
\parbox{0.1\textwidth}{\centering\includegraphics[scale=0.78]{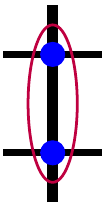}}
$\to$
\parbox{0.1\textwidth}{\centering\includegraphics[scale=0.78]{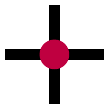}} 
\caption{Example of a $4 \times 4$ two-dimensional tensor network. The circles indicate successive contractions. In every step, the lattice size is reduced by a factor of two until the system contains only a single point.
}
\label{fighotrg}
\end{figure}

\section{Tensor method}
\label{Sec:Method}

Once the model has been formulated in terms of a $d$-dimensional tensor network, the partition function \eqref{Zm} can be computed by summing over all its indices. Thermodynamic observables, which are derivatives of $\ln Z$ with respect to model parameters, can either be computed by taking numerical finite differences of $\ln Z$, or alternatively, by taking analytic derivatives of \eqref{Zm}, which also lead to fully contracted tensor networks, now involving so-called impurity tensors.
Contractions of such tensor networks can be carried out using the HOTRG, which was proposed by Xie et al.\ in 2012 \cite{Xie_2012}. The number of tensors is iteratively reduced by a factor of two by contracting pairs of adjacent tensors and truncating the emerging higher order tensors based on higher order singular value decompositions (HOSVD) \cite{DeLathauwer2000}. 
In Fig.~\ref{fighotrg} we show an example of such a network on a two-dimensional $4 \times 4$ lattice. At each coarsening level, the circles illustrate the contractions of two tensors $T^{(i)}$ at level $i$ to a coarser grid tensor $T^{(i+1)}$ at level $i+1$. As we step from left to right in the figure, the lattice gets coarsened by a factor of two at each step until only a single tensor is left. As a last step we contract the remaining open indices to obtain the partition function. As each direction is contracted in turn, the depicted blocking procedure uses an alternating contraction order. 

In practice we use an improved blocking strategy, which we call improved contraction order (ICO), where, at each blocking step, the contraction direction which yields the smallest (approximate) HOSVD truncation error is chosen \cite{Bloch:2021mjw,Bloch2021}.

The computational cost of the standard HOTRG method scales as $D^{11}$ in three dimensions and as $D^{15}$ in four dimensions.
To improve the computational cost and memory efficiency of the HOTRG method in the four-dimensional case, we introduce an additional approximation where the tensor $T_{tt'xx'yy'zz'}$ is factorized as
\begin{align}
T_{tt'xx'yy'zz'}\approx \sum_{a,b,c,d,e}B^{(t)}_{tt'a}B^{(x)}_{xx'b}B^{(y)}_{yy'c}B^{(z)}_{zz'd} C^{(tx)}_{abe} C^{(yz)}_{cde} .
\label{HT-approx}
\end{align}
If $a,b,c,d,e$ are chosen to be of order $D$, this factorization can be used to reduce the computational complexity of the four-dimensional HOTRG method from $D^{15}$ to $D^{8}$. We call this new method the hierarchical-tensor HOTRG (HT-HOTRG) method.

\section{Results}

\subsection{Chiral condensate}

As we are primarily interested to study dynamical chiral symmetry breaking, the observable of choice is the chiral condensate defined as
\begin{align}
\braket{\bar\psi\psi}
= \frac{1}{V} \frac{\partial\ln Z}{\partial m} .
\end{align}
Our tensor-network results for this observable, which we present below, are computed using the impurity method.

\subsection{Three dimensions}

\begin{figure}
\includegraphics[width=0.49\textwidth]{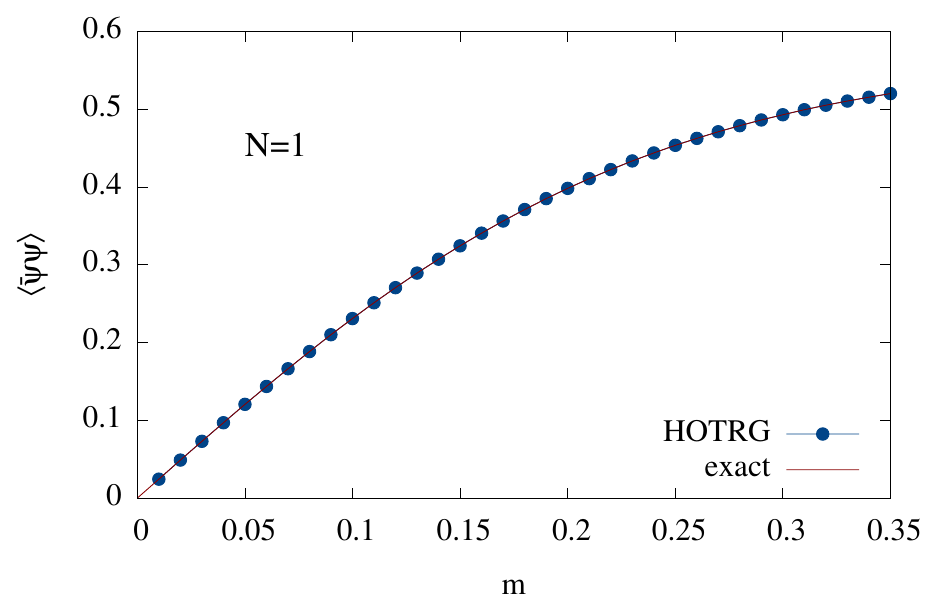}
\hfill
\includegraphics[width=0.49\textwidth]{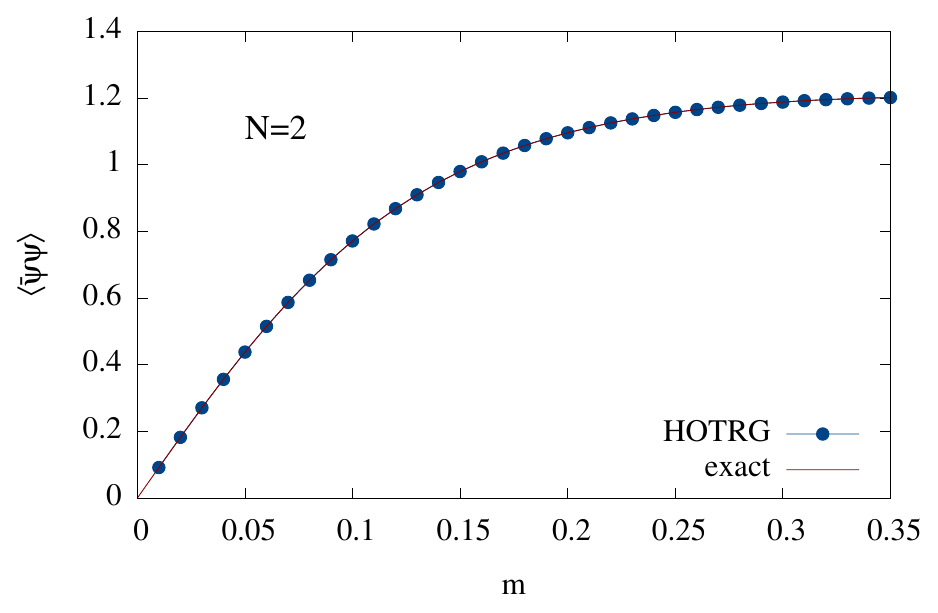}
\caption{Chiral condensate $\braket{\bar{\psi}\psi}$ versus mass $m$ for a three-dimensional $2^3$ lattice. The HOTRG results ($D=12$) agree well with the exact results for $U(1)$ (left) and $U(2)$ (right) \cite{Adams:2003cca}.}
\label{3d-shai}
\end{figure}

We first verify the accuracy of the HOTRG method on a small three-dimensional lattice of volume $2^3$, by looking at the mass dependence of the chiral condensate for $N=1$ and $N=2$, for which exact analytical results have been computed in Ref.~\cite{Adams:2003cca}. 
The numerical HOTRG results in Fig.~\ref{3d-shai}, which show the rise of the chiral condensate with the mass, match the analytic predictions up to twelve digits. 

\begin{figure}
\centerline{\includegraphics[width=0.5\textwidth]{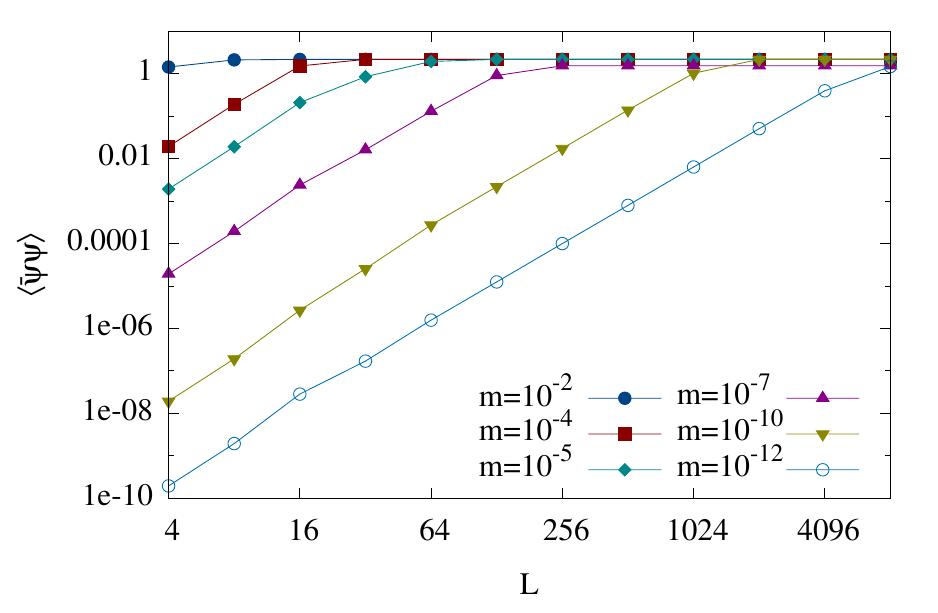}}
\caption{Chiral condensate $\braket{\bar\psi\psi}$ versus lattice size $L$ for $U(3)$ computed using HOTRG with $D=12$. Dynamical chiral symmetry breaking is observed as the chiral condensate convergences to a constant when the lattice size reaches a threshold $L_{\rm th}(m)$, which grows with decreasing mass.
}
\label{3d-finitevol}
\end{figure}

One of the big advantages of tensor methods is the possibility to simulate large lattice volumes at logarithmic cost only. This allows us to investigate spontaneous symmetry breaking, where the chiral condensate is non-zero in the limit that the mass goes to zero while the volume is taken to infinity. This property is related to the correlation length of the system and can also be observed for small masses when the lattice size is sufficiently large.
This is illustrated in Fig.~\ref{3d-finitevol}, where we show the chiral condensate for $U(3)$ as a function of the lattice extent $L$ measured using HOTRG with $D=12$ for different masses $m$ ranging from $10^{-12}$ to $10^{-2}$. For small masses we see that the chiral condensate is affected by finite volume effects when $L$ is too small (smaller than the correlation length) but that it converges for larger volumes. The results from Fig.\ \ref{3d-finitevol} are consistent with $\lim_{m \to 0}\lim_{V \to \infty}\braket{\bar{\psi}\psi} \ne 0$, i.e., even without an explicit symmetry breaking mass term, the chiral symmetry is broken in the strongly coupled $U(N)$ theory. 

\begin{figure}
\centering
\begin{minipage}{0.6\textwidth}
\centerline{\includegraphics[width=0.8\linewidth]{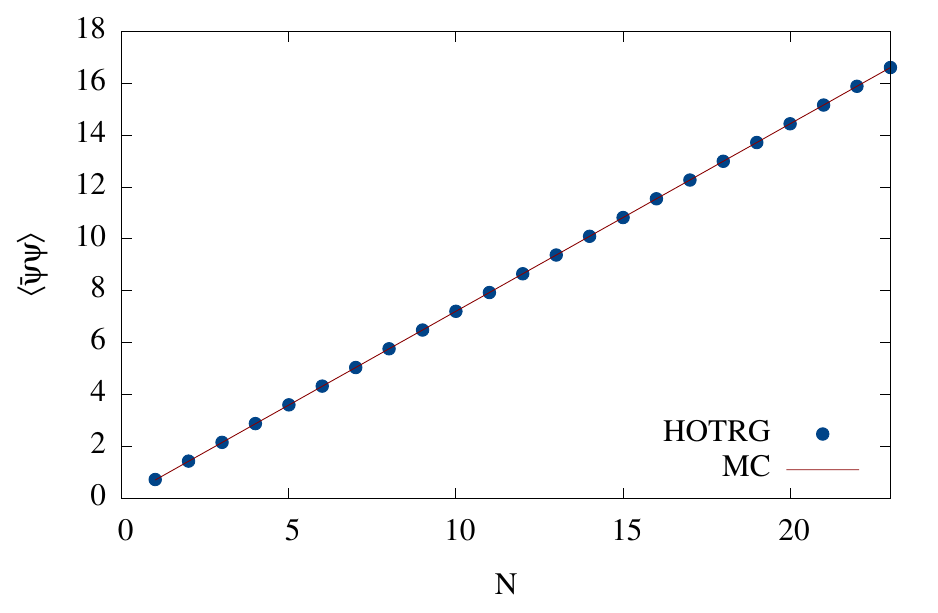}}
\end{minipage}
\begin{minipage}{0.35\linewidth}
\footnotesize
  \begin{tabular}{|c|l|}
  	\hline
    & $\lim\limits_{N \to \infty}\dfrac{\braket{\bar{\psi}\psi}}{N}$ \\
    \hline
   \text{HOTRG ($D=12$)} &  0.7218(4) \\
   	\hline
   \text{Metropolis }  &  0.72311(4)  \\
    \hline
 \end{tabular}
\end{minipage}
\caption{Chiral condensate $\braket{\bar\psi\psi}$ versus degree $N$ of the $U(N)$ group on a three-dimensional $16^3$ lattice for $m=0.1$. The chiral condensate rises linearly with $N$. The slopes $\lim_{N \to \infty}\frac{\braket{\bar{\psi}\psi}}{N}$ extracted from the fit $\braket{\bar{\psi}\psi}=aN+b$ for both simulations in the plot are given in the table on the right. The error for the Metropolis result is a combined statistical and fitting error. For the HOTRG results only a fitting error is quoted.}
\label{3d-Ndep}
\end{figure}

Next, we show the dependence of the chiral condensate on the degree $N$ and discuss its limit as $N \to \infty$. 
To validate the HOTRG results, we compare them with Metropolis Monte Carlo data on a $16^3$ lattice. We perform this comparison for $m=0.1$, as the Metropolis simulations have difficulties to reach lower masses with large enough accuracy.
To leading order, the chiral condensate in Fig.~\ref{3d-Ndep} is linear in $N$, and we therefore make a linear fit $\braket{\bar{\psi}\psi}=aN+b$. We tabulate the fitted slopes, i.e., $\lim_{N \to \infty}\frac{\braket{\bar{\psi}\psi}}{N}$ for both algorithms in the table in Fig.~\ref{3d-Ndep}, and conclude that HOTRG is consistent with Metropolis with a deviation of about $0.2\%$.

\begin{figure}
\centerline{\includegraphics[width=0.5\textwidth]{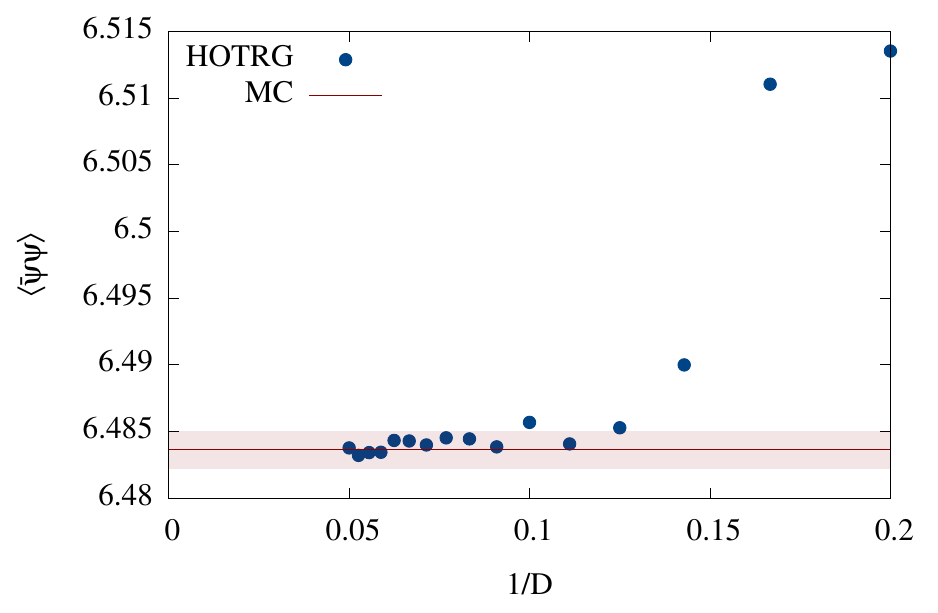}}
\caption{Dependence of $\braket{\bar{\psi}\psi}$ on the HOTRG bond dimension $D$ for a three-dimensional $16^3$ lattice with $m=0.1$ and $N=9$, plotted as a function of $1/D$. The solid line gives the Metropolis result with its error band.  }
\label{3d-Ddep}
\end{figure}

Finally, we investigate the effect of the HOTRG bond dimension $D$ on the results. This parameter is specific to the tensor method and not a $U(N)$ model parameter. In principle one would recover the correct results when taking $D\to \infty$. However, as the memory and computational costs rapidly increase with the bond dimension, the aim is to find a suitable $D$, such that the results will be accurate enough without the computation becoming prohibitively expensive. In Fig.~\ref{3d-Ddep}, the $D$-dependence of $\braket{\bar{\psi}\psi}$ is plotted alongside a Metropolis benchmark for a $16^3$ lattice with $N=9$ and $m=0.1$. For large $D$ the HOTRG results converge to the Metropolis result within its errors.

\subsection{Four dimensions}

\begin{figure}
\centerline{\includegraphics[width=0.5\textwidth]{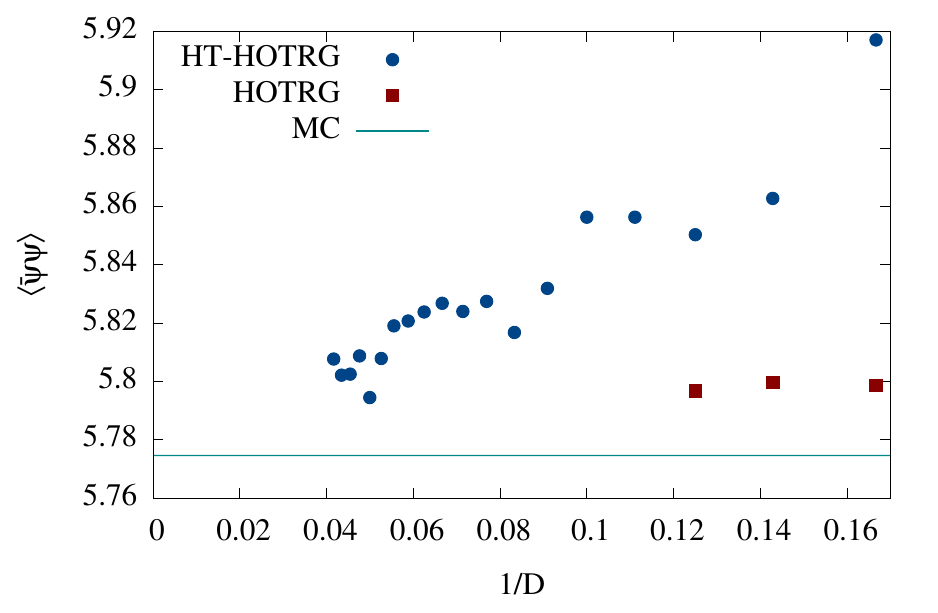}}
\caption{Dependence of $\braket{\bar{\psi}\psi}$ on the bond dimension $D$ for a four-dimensional $16^4$ lattice with $N=9$ and $m=0.1$, plotted as a function of $1/D$. We compare the results obtained with the HT approximation to HOTRG (blue dots), the standard HOTRG (red squares) and the Metropolis method (solid line). For large $D$ the HT approximation is consistent with the HOTRG results. Although the tensor results are close to the MC data, they are clearly not yet converged for these $D$ values.}\label{4d-Ddep}
\end{figure}

A similar analysis can be carried out for four dimensions, except for the comparison with the exact results. In four dimensions, the standard HOTRG algorithm has a memory cost of $D^{8}$ and complexity of $D^{15}$, and therefore, the accessible range of $D$ is very limited. To extend this range, we use the HT-HOTRG method, where the tensor is approximated by the factorization \eqref{HT-approx} throughout the simulation. In Fig.~\ref{4d-Ddep} we compare the results obtained with HT-HOTRG, standard HOTRG and Metropolis MC on a $16^4$ lattice for $N=9$ and $m=0.1$. Even though we find a qualitative agreement between the tensor and the MC results, the former have not yet converged for the $D$ values used in our simulations. We also observe that the results obtained with the HT approximation agree with the standard HOTRG, when $D$ is taken large enough. Therefore, we use HT-HOTRG with $D=20$, as it has the same accuracy as standard HOTRG with $D=9$, but only requires about $0.5\%$ of the computation time.

\begin{figure}
\centerline{\includegraphics[width=0.5\textwidth]{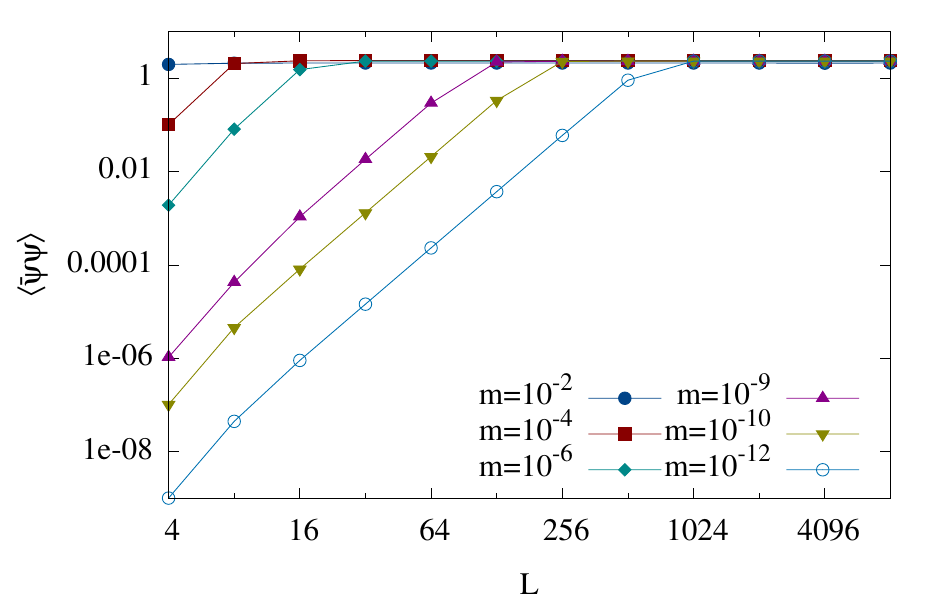}}
\caption{Chiral condensate $\braket{\bar{\psi}\psi}$ as a function of the lattice extent $L$ for varying small masses for $N=3$ and $D=20$.}\label{4d-finitevol}
\end{figure}

We also investigate dynamical chiral symmetry breaking in four dimensions. There are again finite volume effects which obscure the DCSB when the system size becomes smaller than the correlation length for small masses. This can be seen by fixing the mass and varying the linear extent $L$ of the lattice, as is shown in Fig.~\ref{4d-finitevol}. Just like for the three-dimensional case, we observe that the lattice has to be increased with decreasing mass to obtain a signal for dynamical chiral symmetry breaking.

\begin{figure}
\centering
\begin{minipage}{0.6\textwidth}
\centerline{\includegraphics[width=0.8\textwidth]{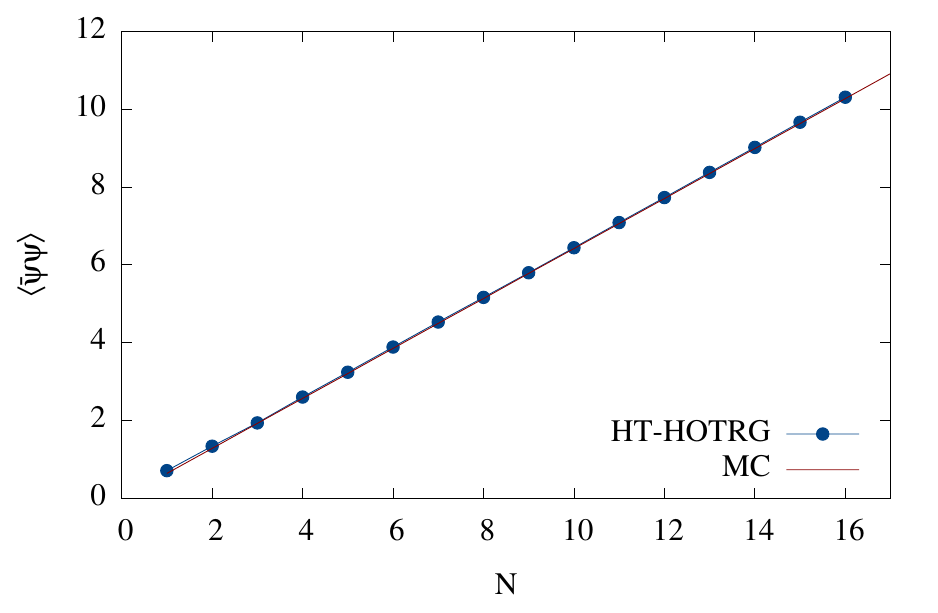}}
\end{minipage}
\begin{minipage}{0.35\linewidth}
\footnotesize
  \begin{center}
  \begin{tabular}{|c|l|}
  	\hline
    & $\lim\limits_{N \to \infty}\dfrac{\braket{\bar{\psi}\psi}}{N}$ \\
    \hline
   \text{HT-HOTRG (D=20)} 	 &  0.637(2) \\
    \hline
   \text{Metropolis}   &  0.64279(1) \\
   	\hline
 \end{tabular}
 \end{center}
\end{minipage}

\caption{Dependence of the chiral condensate on the degree $N$ on a four-dimensional $16^4$ lattice with $m=0.1$, measured using the HT-HOTRG approximation with $D=20$. Since the memory cost for the initial tensor is still very high for large $N$, even with HT-HOTRG, we could only reach $N=16$. The chiral condensate depends linearly on $N$.
The fitted slope is given in the table, where the error for the Metropolis result is a combined statistical and fitting error, while for the HT-HOTRG result we only determine a fitting error.
}\label{4d-Ndep}
\end{figure}

We also investigate the $N$ dependence of the chiral condensate and compare it with the Metropolis results on a $16^4$ lattice with $m=0.1$. This is shown in Fig.~\ref{4d-Ndep}, where we observe good agreement between both methods.
As the chiral condensate is linear in $N$, we also make a linear fit and report the slope in the table included in the  figure. The limit of $\frac{\braket{\bar{\psi}\psi}}{N}$ for $N \to \infty$ differs less than $1\%$ between both methods.

\section{Summary}

We have applied the HOTRG method to the strong-coupling limit of $U(N)$ gauge theory in three and four dimensions and found good agreement with Metropolis Monte Carlo results. In three dimensions we first verified that the tensor method reproduces the exact results on a $2^3$ lattice. The tensor method enabled us to investigate dynamical chiral symmetry breaking in three and four dimensions, as we can easily simulate large lattice volumes, which are required when approaching the zero mass limit. In four dimensions we used an additional HT approximation to HOTRG to reduce the computation time. We also studied the $N$-dependence of the chiral condensate and verified the convergence of the results as a function of the bond dimension.

In the future we plan to apply tensor network methods to strongly coupled $SU(3)$ gauge theory, where a dual representation in terms of dimers, monomers and baryon loops exists \cite{Karsch:1988zx}.

\nocite{apsrev42Control}
\bibliographystyle{apsrev4-2.bst}
\bibliography{biblio,revtex-custom}

\end{document}